\documentclass[a4paper,conference]{IEEEtran}

\usepackage{cite}
\usepackage{float}

\usepackage{graphicx}
 
\usepackage[cmex10]{amsmath}
\usepackage{algorithm}
\usepackage{algorithmic}

\usepackage{array}
\usepackage[font=footnotesize]{subfig}

\begin{document}

\title{Using Correlated Subset Structure for Compressive Sensing Recovery}

\author{\authorblockN{Atul Divekar}
\authorblockA{Alcatel-Lucent\\
Naperville, IL 60563\\
Email: atul.divekar@alcatel-lucent.com}
\and
\authorblockN{Deanna Needell}
\authorblockA{Department of Mathematics\\ 
Claremont McKenna College\\
Claremont, CA 91711\\
Email: dneedell@cmc.edu}
}

\newcommand{\vnorm}[1]{\left|\left|#1\right|\right|}

\maketitle
\begin{abstract}
Compressive sensing is a methodology for the reconstruction of sparse 
or compressible signals using far fewer samples than required by the
Nyquist criterion.  However, many of the results in compressive sensing
concern random sampling matrices such as Gaussian and Bernoulli matrices.
In common physically feasible signal acquisition and reconstruction 
scenarios such as super-resolution of images, the sensing matrix has
a non-random structure with highly correlated columns.
Here we present a compressive sensing recovery algorithm that exploits
this correlation structure.  We provide algorithmic justification as well 
as empirical comparisons.
\end{abstract}

\section{Introduction}
 
  Consider the problem of image super-resolution, where one or more low-resolution images of a scene are used to synthesize a single image of higher resolution. If multiple images are used, they are commonly assumed to be subpixel-shifted and downsampled versions of the original high resolution image that is to be reconstructed~\cite{superres_survey}. Alternatively, super-resolution from a single low resolution image using a dictionary of image patches and compressive sensing recovery has been proposed in \cite{wright_superres}.  
The relationship between the available low resolution and desired high resolution image is commonly modeled by a linear filtering and downsampling operation. Suppose that we wish to reconstruct a size $N\times N$ high resolution image from a lower resolution image, for example of size $\frac{N}{2}\times \frac{N}{2}$, or smaller.  Let $x$ and $y$ represent the vectorized high and low resolution images respectively. We model the formation of $y$ from $x$ by the equation $y=SHx + \eta$ where $\eta$ is the sensor noise, $S$ is a downsampling matrix of size $\frac{N}{2}* \frac{N}{2}$ by $N^{2}$, and $H$ is a $N^{2}$ by $N^{2}$ matrix that represents the filtering (antialiasing) operation. In order to consider super-resolution as a compressive sensing recovery problem we write $x=\Psi c$ where $\Psi$ is a sparsifying basis for the class of images under consideration and $c$ is the coefficient vector corresponding to image $x$ with respect to the basis $\Psi$. 
In the simplest case, $\Psi$ is an $N^{2}\times N^{2}$ orthogonal matrix, but can also be generalized to an overcomplete dictionary.
Here we have 
$
y = SH\Psi c + \eta = \Phi c + \eta, 
$
where $\Phi=SH\Psi$ is the sampling matrix. 

 Most of the work in the compressive sensing literature assumes $\Phi$ to be random matrix, such as a partial DFT or one drawn from a Gaussian or Bernoulli distribution. However, in this scenario
the matrix is not random, but instead has correlated columns whose structure we wish to exploit to improve compressive sensing recovery. Here we assume that $H$ is not a perfect low pass filter, so that it is possible for $\Phi=SH\Psi$ to preserve enough high frequency information for recovery to be possible; $SH$ and $\Psi$ have sufficient incoherency to allow $c$ to be recovered with acceptable error.

% We consider here superresolution from a single low resolution image using compressive sensing with a more direct approach, and use it to develop a novel compressive sensing recovery algorithm.

 Compressed sensing provides techniques for stable sparse recovery~\cite{candes2006robust,candes2006stable,donoho1989uncertainty}, but results for coherent sensing matrices have been limited~\cite{candes2011compressed,candes2012towards,fannjiang2012coherence}.
 
{\bf Organization.} The structure we wish to exploit is first described. Then we present algorithms that take advantage of this structure for compressive sensing recovery.

\section{Correlation Structure}

Typical examples of sparsifying bases $\Psi$ for images are wavelets and blockwise discrete cosine transform bases. Images exhibit correlation at each scale: neighboring pixels are heavily correlated except across edges, local averages of neighboring blocks are heavily correlated except across edges, and so on. This makes wavelet-like bases, which have locally restricted atoms, suitable for sparsifying the image. For the super-resolution setting with the low resolution image of size $\frac{N}{2}\times \frac{N}{2}$, the rows of $SH$ consist of shifted versions of the filtering kernel with shifts of 2 horizontally and vertically. Due to the localized nature of wavelet bases, we expect columns of $\Phi$ that correspond to spatially distant bases in $\Psi$ to have little correlation. If $\Psi$ is a tree structured orthogonal wavelet basis matrix, columns of $\Psi$ that overlap spatially are orthogonal, however when filtered by $H$, they result in significant correlation. Then we expect columns in $\Phi$ to show significant correlation in tree structured patterns.

We illustrate this with an example. For simplicity we consider only one-dimensional signals,
though the discussion is equally valid for images. Suppose that $\Psi$ is a $256\times 256$ matrix whose columns consist of the length 256 Haar basis vectors, and $SH$ is a $128\times 256$ matrix obtained by shifting the filter kernel $h=\{0.1,0.2,0.4,0.2,0.1\}$ by two from one row to the next. $SH$ represents the filtering and downsampling operation that generates the low resolution signal $y=SHx$ from the length 256 signal $x$. Then $\Phi=SH\Psi$ is the sampling matrix. 

\begin{figure}[h]
\vspace{-1mm}
\centering
  \includegraphics[height=4.0cm,width=4.0cm]{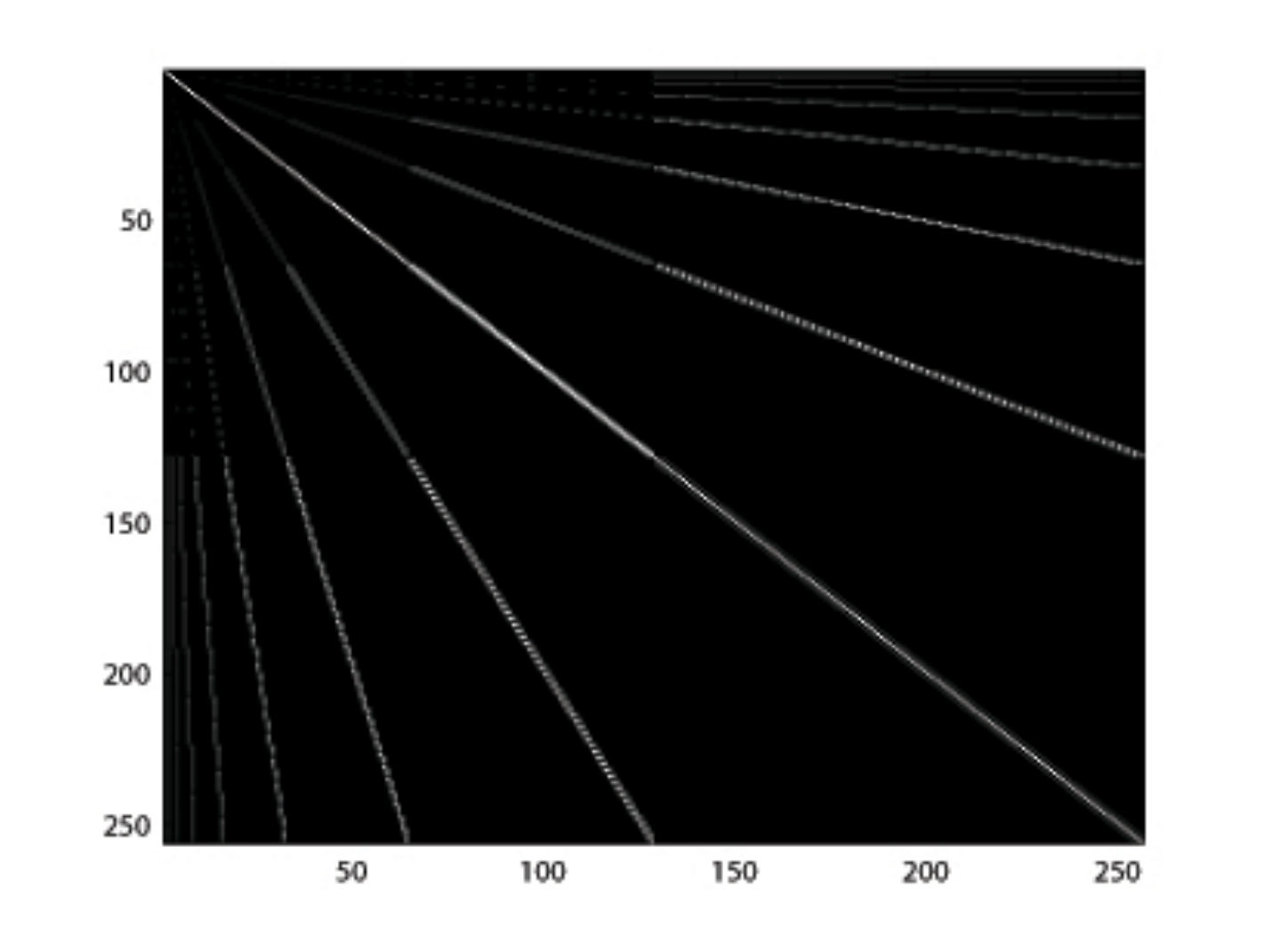} 
  \vspace{-1mm}
  \caption{Absolute values of $\Phi^{*}\Phi$.\label{corr_struct}}
  \vspace{-1mm}
\end{figure}

Fig. \ref{corr_struct} shows the absolute values of the correlation matrix $C=\Phi^{*}\Phi$ (here and throughout $A^*$ denotes the adjoint of $A$). This shows that only a small number of pairs of columns of $\Phi$ are strongly correlated to each other. Each filtered wavelet basis is correlated with other spatially overlapping bases at coarser and finer scale and in the immediate neighborhood, but has no correlation with spatially distant bases.

More generally, consider compressive sensing recovery where the columns of the sampling matrix $\Phi$ can be grouped into nearly-isolated sets, such that correlation among pairs of columns within a set may be significant, but correlation between two columns that belong to different sets is relatively small. How does one exploit this structure to efficiently reconstruct the signal?  

One of the central results in compressive sensing is that if matrix $\Phi$ exhibits a property called the Restricted Isometry Property (RIP)~\cite{decode_tao,RV08:sparse}, convex optimization can recover the sparse signal 
exactly~\cite{candes_tao,candes} via %Briefly, for a matrix $\Phi$ the $r^{th}$ restricted isometry constant is the 
%least constant $\delta_{r}$ for which
%
%\begin{equation} \label{RIP}
%(1-\delta)||c||_{2}^{2} \leq || \Phi c||^{2} \leq (1+\delta)||c||_{2}^{2}
%\end{equation}
%
%whenever $||c||_{0} := |\supp(c)| \leq r$.
%Matrices with elements drawn from a Gaussian or Bernoulli distribution are known to satisfy RIP with overwhelming probability~\cite{RV08:sparse,BDDW07:Johnson-Lindenstrauss}.
%
%Then it is shown in \cite{candes_tao,candes}, for example, that if $\Phi$ satisfies the RIP with $r=2K$ and $\delta_{r} < \sqrt{2}-1$, then $c$ can be recovered perfectly by solving

\begin{equation} \label{eq:l1}
\text{min } \vnorm{c}_{1} 
 \text{such that } y= \Phi c.
\end{equation}

However, the sampling matrix $\Phi=H\Psi$ described above does not obey the RIP and these results are not readily applicable. On the other hand, it is commonly found in practical applications and has a structure that could be exploited.

Before considering the above problem, a simple modification to CoSaMP\cite{CoSaMP} is presented that provides some improvement in recovery performance. This algorithm, called
Partial Inversion (PartInv) and described by Algorithm~\ref{PartInv}, also indicates how the above described structure could be exploited.

\section{Partial Inversion}

Consider the usual CS setting: Given a length $M$ sample vector $y=\Phi c + \eta$ where $\Phi$ is an $M\times N$ sampling matrix and $c$ a length $N$ vector with sparsity $K < M$, we wish to obtain the best $K$-sparse approximation $\hat{c}$ to $c$. At each step let $I$ be an index set, so
that for example, $\hat{c}_{I}$ represents an estimate of the components of $c$ corresponding to the column indices in $I$. $\hat{c}$ by itself is an estimate for all the columns 
$\{1..N\}$. Let $L$ for $K\leq L<M$ be an adjustable parameter for the size of the 
set $I$. We get good results with $L=\max\{K,0.8M\}$. Let $\Phi_{I}$ denote the matrix of columns from $\Phi$ corresponding to indices in the set $I$. Let $\tilde{I}=\{1..N\} \backslash I $ denote the complement of $I$. For any full rank matrix $A$, define $A^{\dagger}=(A^{*}A)^{-1}A^{*}$.

\begin{algorithm}[htbp]
%\vspace{-1mm}
\caption{Given $y=\Phi c$, return best $K$-sparse approximation $\hat{c}$ }
\label{PartInv}
\begin{algorithmic}[1]
\STATE $\hat{c} \leftarrow \Phi^{*}y; I^{0} \leftarrow$ indices of the $L$-largest magnitudes of $\hat{c} ;  k \leftarrow 0$
\WHILE{Stopping condition not met} 
\STATE $\hat{c}_{I^{(k)}} \leftarrow \Phi_{I^{(k)}}^{\dagger} y$
\STATE $r \leftarrow y - \Phi_{I^{(k)}}\hat{c}_{I^{(k)}}$
\STATE $J^{(k)} \leftarrow \widetilde{I^{(k)}}$ 
%\STATE $P \leftarrow [Id-\Phi_{I^{(k)}} \Phi_{I^{(k)}}^{\dagger}]\Phi_{J^{(k)}}$
%\STATE $\hat{c}_{J^{(k)}} \leftarrow P^{*}r$
\STATE $\hat{c}_{J^{(k)}} \leftarrow \Phi_{J^{(k)}}^{*}r$
\STATE $I^{(k+1)} \leftarrow$ indices of the $L$-largest magnitude components of $\hat{c}$.
\STATE $ k \leftarrow k+1$
\ENDWHILE
\end{algorithmic}
%\vspace{-1mm}
\end{algorithm}

For the noiseless case $\eta=0$, the stopping condition can be obtained by testing the 
magnitude of $r_{2}= y-\Phi \hat{c}$ at the start of each iteration.  If set $I$ does not
vary from one iteration to the next, the algorithm cannot progress further and can be 
stopped immediately.  In practice the inversion of line 3 can be done efficiently by Richardson's algorithm (see e.g. Sec. 7.2 of~\cite{Bjo96:Numerical-Methods}).  %If $|J^{(k)}|<M$, we may replace line 6 by $\hat{c}_{J^{(k)}} \leftarrow \Phi_{J^{(k)}}^{\dagger}r$. 

%
%\begin{algorithm}[htbp]
%\label{algo:Richardson}
%\caption{Richardson's iteration: Given full rank matrix $\Phi_{I}$ returns $\hat{c}_{I}= \Phi_{I}^{\dagger} y$}
%\begin{algorithmic}[1]
%\STATE $R \leftarrow \Phi_{I}^{*}\Phi_{I}-I$
%\STATE $\hat{c}_{I}^{(0)} \leftarrow 0$
%\STATE $w \leftarrow \Phi_{I}^{*}y$
%\WHILE{Stopping condition not reached}
%\STATE $\hat{c}_{I}^{(n+1)} \leftarrow w - R \hat{c}_{I}^{(n)}$
%\ENDWHILE
%\end{algorithmic}
%\end{algorithm}

This algorithm demonstrates improvement relative to CoSaMP when the accurate recovery region
is considered on a plot of $\frac{K}{M}$ versus $\frac{M}{N}$. The motivation is the following (for simplicity we drop the iteration indicator $k$) :
From line 3, 

\begin{align}
\label{eqn:PartInv}
\hat{c}_{I} &= \Phi_{I}^{\dagger} y\\
            &= c_{I} + (\Phi_{I}^{*}\Phi_{I})^{-1}\Phi_{I}^{*}\Phi_{\tilde{I}}c_{\tilde{I}}.
\end{align}

Compare this to the estimator $\hat{c_{I}} = \Phi_{I}^{*}r$ used in CoSaMP. When $r=y$,
we have 

\begin{align}
\hat{c_{I}} &= \Phi_{I}^{*}y \\
            &= \Phi_{I}^{*}\Phi_{I}c_{I} + \Phi_{I}^{*}\Phi_{\tilde{I}}c_{\tilde{I}} \\
            &= c_{I} + (\Phi_{I}^{*}\Phi_{I}-I)c_{I} + \Phi_{I}^{*}\Phi_{\tilde{I}}c_{\tilde{I}}
\end{align}

If the index set $I$ contains several nonzero coefficients
(which we hope is true),  then $(\Phi_{I}^{*}\Phi_{I}-I)c_{I}$, which results from the mutual interference between the columns of $\Phi_{I}$, is significant and is a source of noise in $\hat{c_{I}}$. This term is eliminated in~\eqref{eqn:PartInv}. Partial inversion does add $(\Phi_{I}^{*}\Phi_{I})^{-1}$ to the remaining noise term, however, the singular values of this term can be kept from significantly amplifying the noise term by a conservative choice of $L$, the size of the index set $I$ (for example, empirically we find that $L=s$ tends to be a safe choice, but larger values often lead to noise amplification for certain types of matrices).
The improved estimate $\hat{c}_{I}$ further produces an improved estimate $\hat{c}_{J^{(k)}}$, which leads to a better selection of nonzero coefficients in the next iteration.

 The expression~\eqref{eqn:PartInv} also indicates how the correlation structure 
may be used to improve recovery. The noise term $(\Phi_{I}^{*}\Phi_{I})^{-1}\Phi_{I}^{*}\Phi_{\tilde{I}}c_{\tilde{I}}$ depends upon the correlation between the sets $\Phi_{I}$ and $\Phi_{\tilde{I}}$ given by $\Phi_{I}^{*}\Phi_{\tilde{I}}$. This correlation is weak if $\Phi_{I}$ and $\Phi_{\tilde{I}}$ are sufficiently spread.% out in $R^{M}$. 
 However, the correlation is likely to remain large if $L$ is significant compared to $M$, as will be the case when $\frac{K}{M}$ is large.

\section{Experimental Comparison}

 We compare the recovery performance of Partial Inversion with CoSaMP and convex optimization~\eqref{eq:l1} for two classes of matrices: Gaussian random matrices, and matrices constructed to have highly correlated subsets of columns with low correlation across subsets. 

\begin{figure}
\vspace{-1mm}
   \centering
   \begin{tabular}{@{\hspace{-2mm}}c@{\hspace{-3mm}}c}
   (a) \includegraphics[width=1.6in]{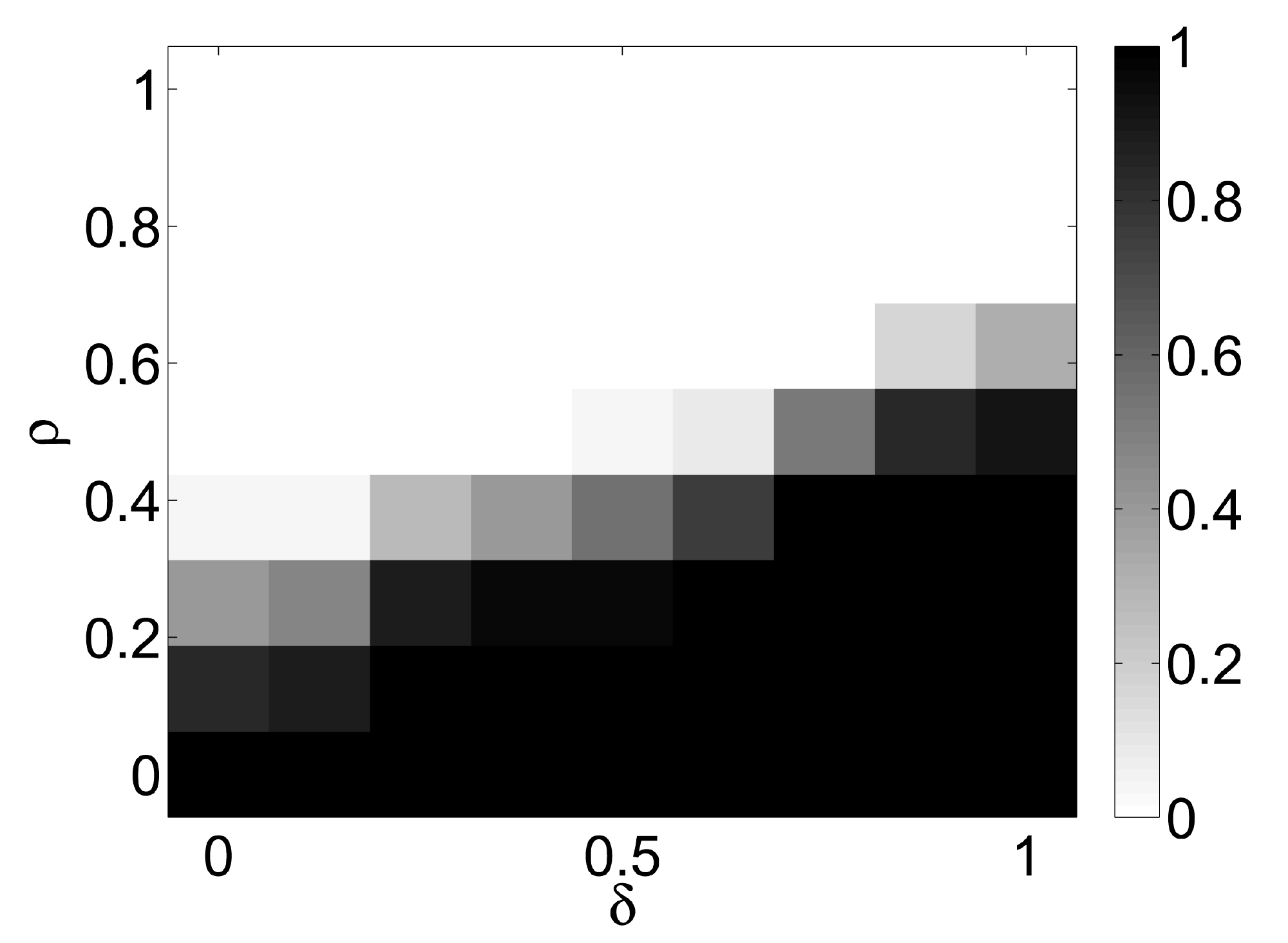} & (d) \includegraphics[width=1.6in]{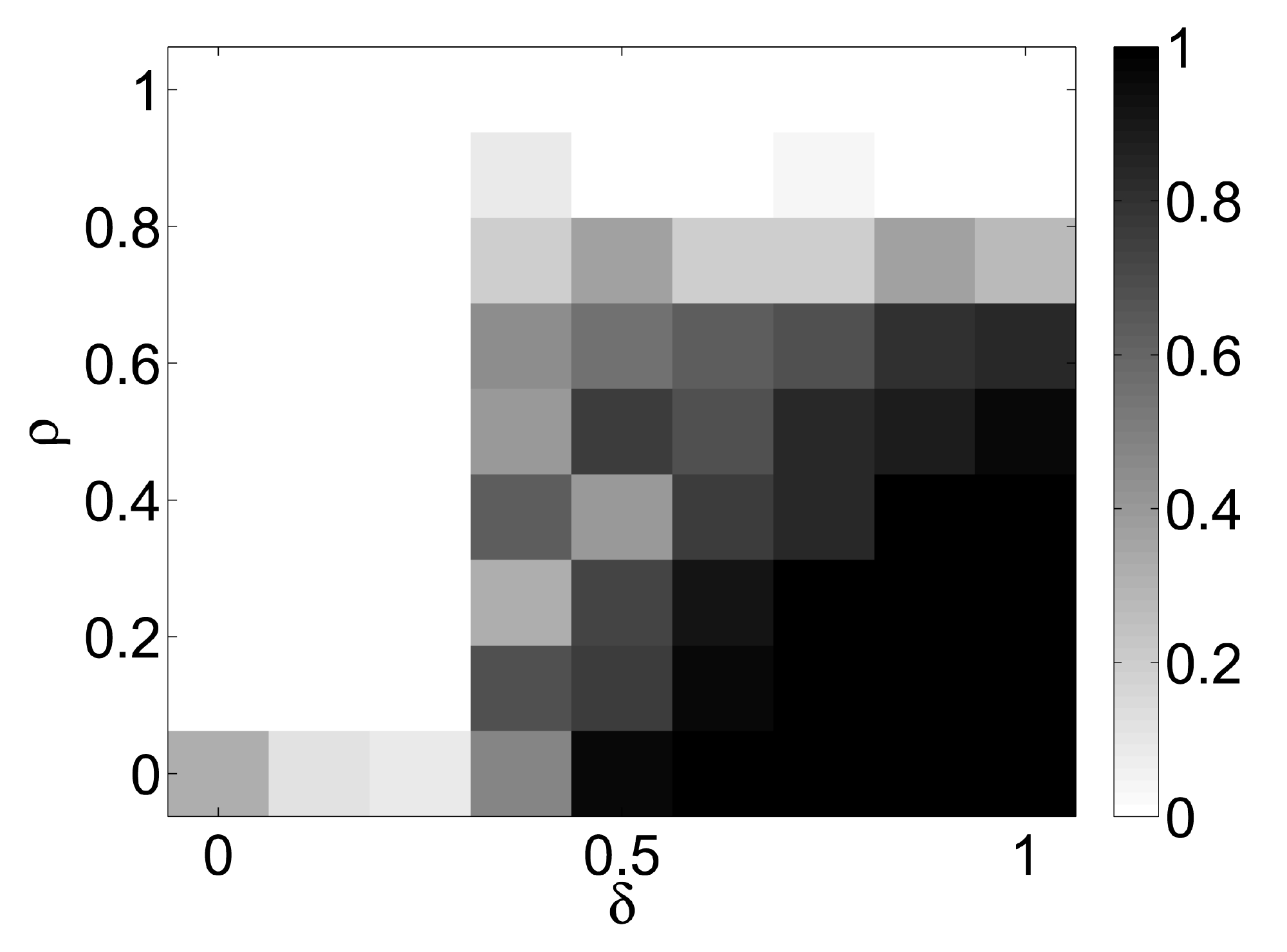}\\
    (b) \includegraphics[width=1.6in]{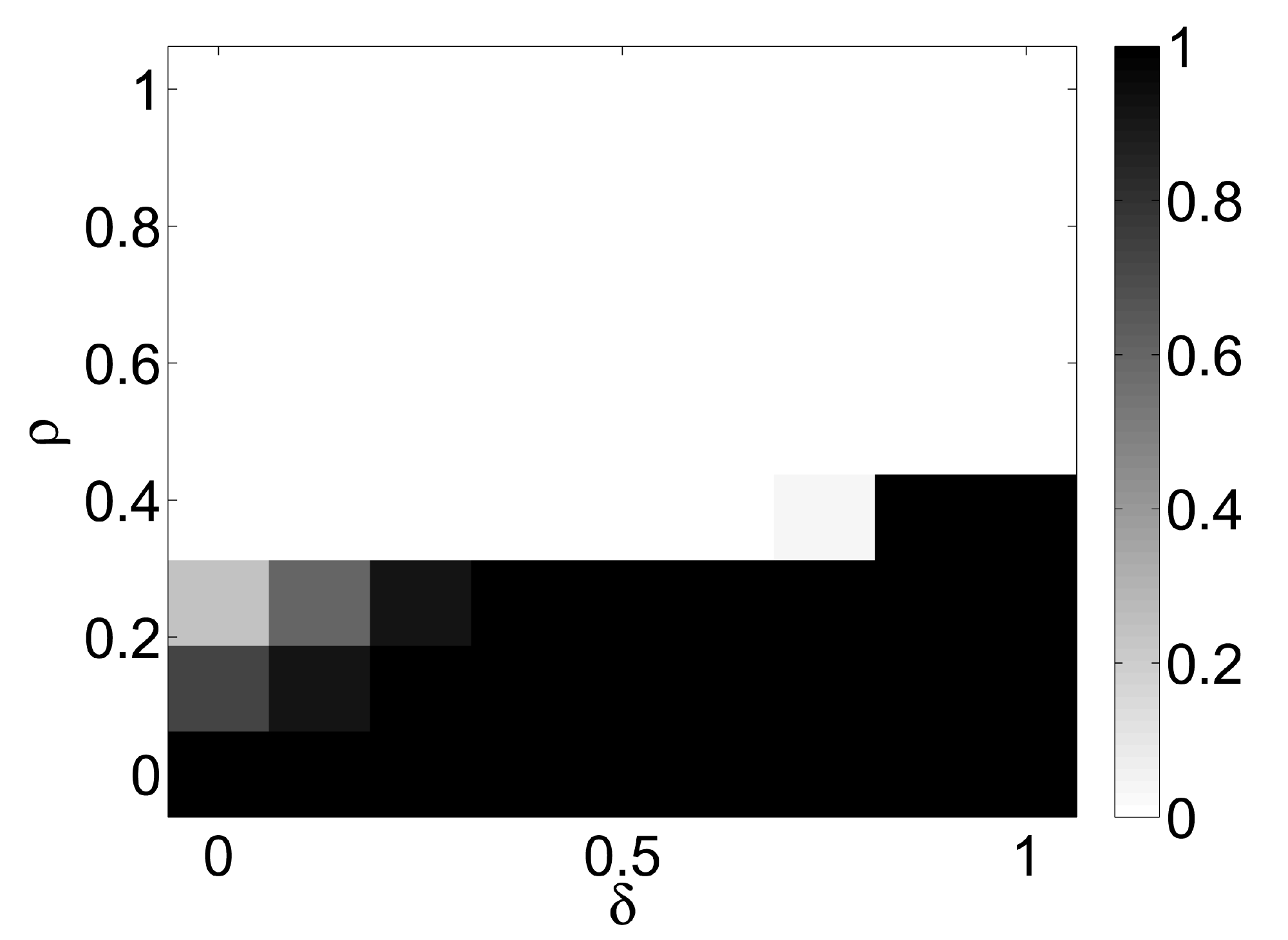} & (e) \includegraphics[width=1.6in]{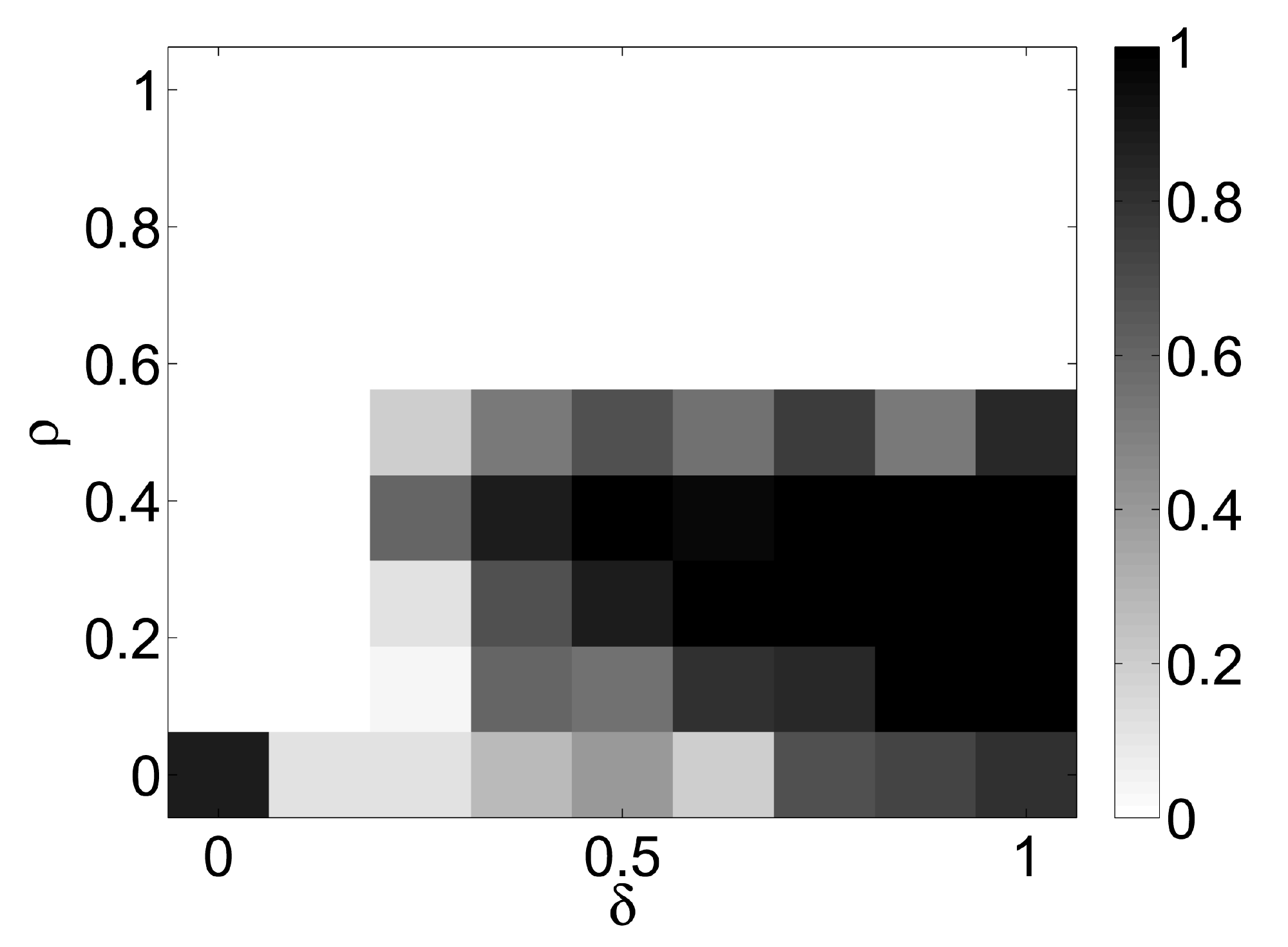} \\
    (c) \includegraphics[width=1.6in]{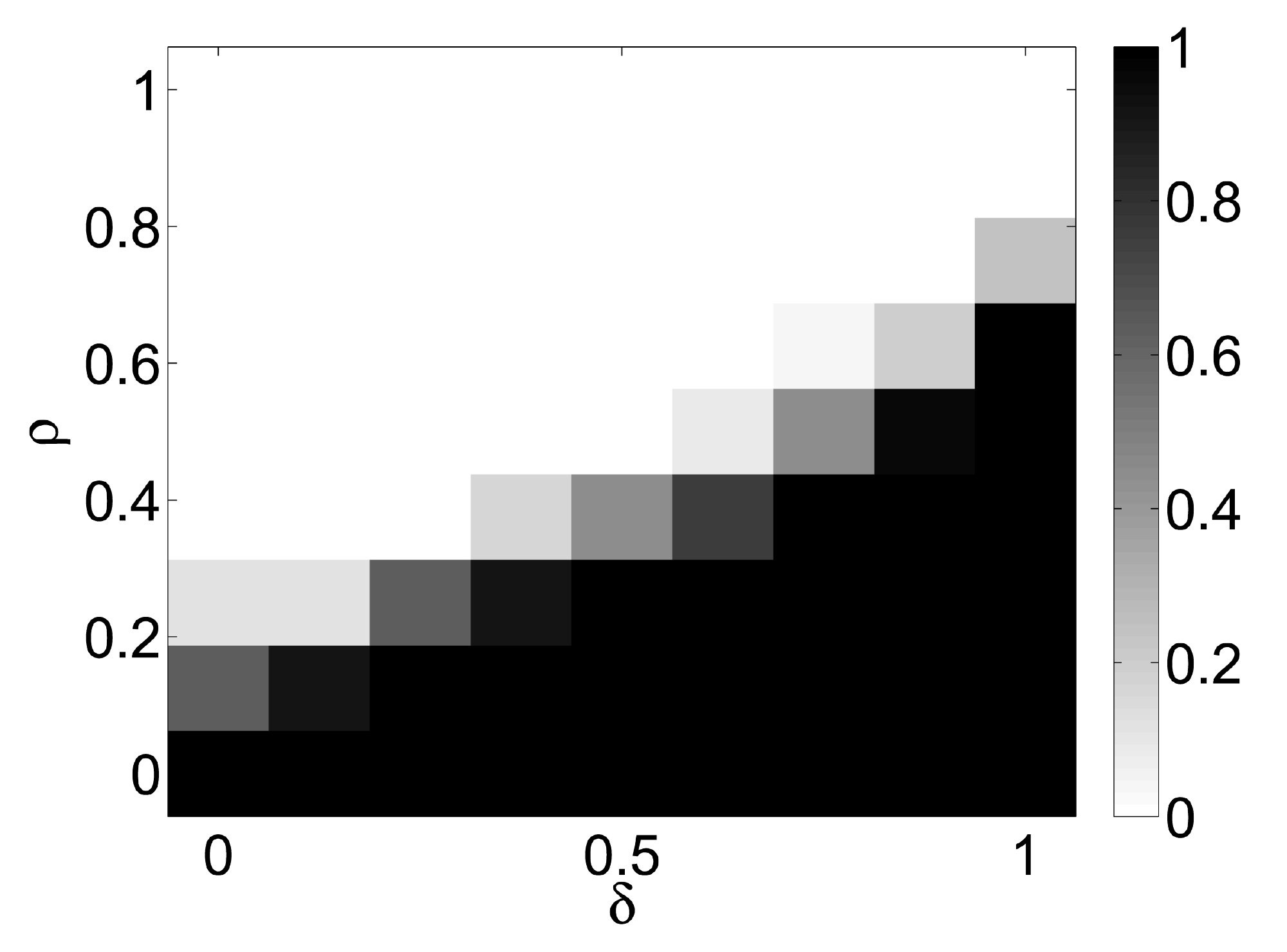} & (f) \includegraphics[width=1.6in]{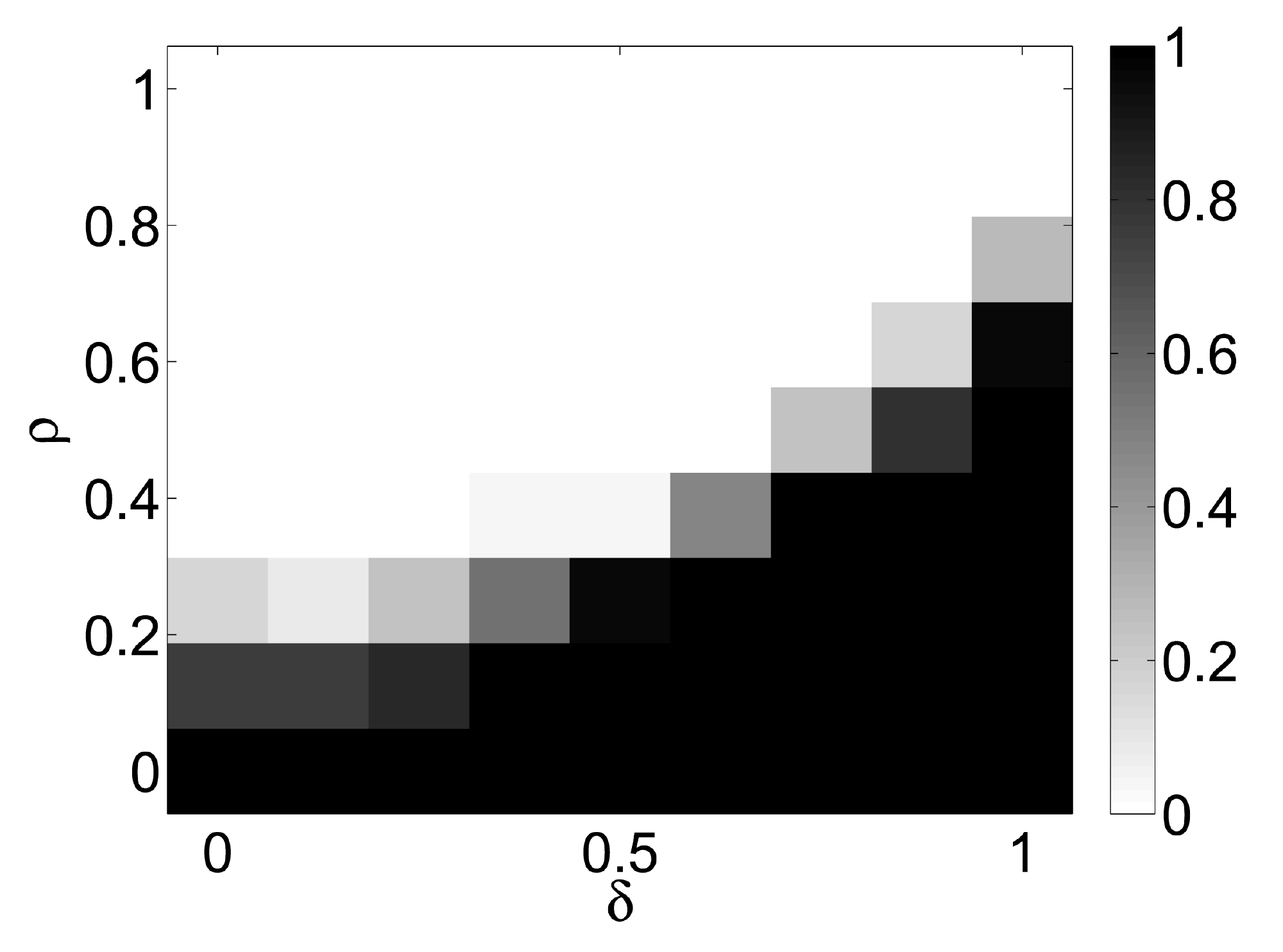} 
  % \hspace*{9mm} (a) & \hspace*{9mm} (b)
   \end{tabular}
   \vspace{-1mm}
   \caption{\small \sl Proportion of successes on Gaussian matrices using (a) PartInv, (b) CoSaMP and (c) $\ell_1$-minimization, and proportion of successes on correlated column subset matrices using (d) PartInv, (e) CoSaMP and (f) $\ell_1$-minimization for various values of $\delta=\frac{M}{N} \in (0,1)$ (horizontal axis) and $\rho=\frac{K}{M} \in (0,1)$ (vertical axis).
   \label{fig:gauss}}
   \vspace{-1mm}
\end{figure}

In the first case, we construct $M$ by $N$ matrices with $N(0,1)$ elements along with the coefficient vector $c$ containing $K$ nonzero entries taken from a $N(0,1)$ distribution. The nonzero locations are selected uniformly at random from $\{1...N\}$. Each column in each matrix is normalized to have unit $l_{2}$ norm. We set $N=256$ and vary $\delta=\frac{M}{N}$ from 0.1 to 0.9 in steps of 0.1. For each $\delta$ we vary $\rho=\frac{K}{M}$ from 0.1 to 0.9 in steps of 0.1. For each $(\delta,\rho)$ point we carry out 25 trials, and declare success if $\frac{1}{N}||c-\hat{c}||^{2}< 10^{-5}$. For PartInv we considered two cases for the size of subset $I$ : $L=S$ and $L=\max\{S,0.8M\}$. We see better performance in the $L=S$ case.  For $l_{1}$ minimization we use the $l_{1}$-magic package \cite{l1_magic}. We show the results in Fig. \ref{fig:gauss}.

In the second case, we construct $M$ by $N$ matrices with $N=256$ and variable $M$ and a block diagonal structure. The columns are divided into 16 column subsets. In each subset we set $M/16$ rows to 1. In addition, to every element of the matrix we add noise drawn from a zero-mean normal distribution with variance 0.0025. This produces heavy intra-subset correlation and light correlation across subsets. We let the coefficient vector $c$ contain
$S$ nonzeros elements drawn from a $N(0,1)$ distribution. We select 4 of the 16 subsets at random and in each subset select $\frac{S}{4}$ of the indices to have nonzero values, again uniformly at random. If some of the nonzeros were left over, they are accomodated in a fifth subset. For PartInv we set $L=\max\{S,0.8M\}$.
The results are also depicted in Fig.~\ref{fig:gauss}.

\section{Recovery of Coefficients Concentrated on Wavelet Trees}

We next use Partial Inversion to recover nonzero coefficients that are concentrated on wavelet 
trees, which is commonly seen when a signal or image with discontinuities is decomposed in a wavelet basis. When the coefficients are concentrated on an isolated set (a set of columns that have low correlation with columns outside the set), a setwise estimator is especially useful
to identify the sets on which the coefficients are nonzero. Consider the 2D wavelet case.
Suppose that $I$ is the index set of columns of the wavelet basis belonging to a particular tree rooted at a coarse scale and containing finer scale coefficients. We have

\begin{equation} 
\label{eq:z}
z_{I} = \Phi_{I}^{*}y \\
      = \Phi_{I}^{*}\Phi_{I}c_{I} + \Phi_{I}^{*}\Phi_{\tilde{I}}c_{\tilde{I}}.
\end{equation}

Because $\Phi_{I}$ is relatively isolated from the columns in $\Phi_{\tilde{I}}$, the second term is small, and because most of the elements of $c_{I}$ are nonzero, the first term is large.
This is further intensified by the mutual correlation of the columns of $\Phi_{I}$ which is high because of the spatial overlap of the support of the wavelet bases in the tree. This motivates a simple selection criterion for measuring the strength of the nonzero coefficients in each wavelet tree $I$: $s_{I}=\sum\limits_{j\in I}|z_{j}|$. We use this criterion along with PartInv to select wavelet trees that are known to be nonzero. We denote the number of subsets by SETNUM.

We modify the PartInv algorithm to use this estimator.

\begin{algorithm}[htbp]
%\vspace{-1mm}
\caption{Given $y=\Phi c$, with $K$ nonzero coefficients concentrated on wavelet trees,return best $K$-sparse approximation $\hat{c}$ }
\label{PartInv-Wavelet}
\begin{algorithmic}[1]
\STATE $\hat{c} \leftarrow \Phi^{*}y$; 
\STATE $k \leftarrow -1$
\FOR{$j=1 \to \text{SETNUM}$}
\STATE $s_{j} \leftarrow \sum\limits_{l\in I_{j}}|\hat{c}_{l}|$
\ENDFOR
\STATE $I^{k+1} \leftarrow$ indices of columns contained in the sets with the largest magnitude $s_{k}$, to include at least $K$ coefficients.
\STATE $ k \leftarrow k+1$
\WHILE{Stopping condition not met} 
\STATE $\hat{c}_{I^{(k)}} \leftarrow \Phi_{I^{(k)}}^{\dagger} y$
\STATE $r \leftarrow y - \Phi_{I^{(k)}}\hat{c}_{I^{(k)}}$
\STATE $J^{(k)} \leftarrow \widetilde{I^{(k)}}$ 
%\STATE $P \leftarrow [Id-\Phi_{I^{(k)}} \Phi_{I^{(k)}}^{\dagger}]\Phi_{J^{(k)}}$
%\STATE $\hat{c}_{J^{(k)}} \leftarrow P^{*}r$
\STATE $\hat{c}_{J^{(k)}} \leftarrow \Phi_{J^{(k)}}^{*}r$
\STATE Repeat lines $3-6$
%\FOR{$j=1 \to SETNUM$}
%\STATE $s_{j} \leftarrow \sum\limits_{l\in I_{j}}|\hat{c}_{l}|$
%\ENDFOR
%\STATE Repeat line $6$%$I^{(k+1)} \leftarrow$ indices of columns contained in the sets with the largest magnitude $s_{k}$,to include atleast $K$ coefficients.
\STATE $ k \leftarrow k+1$
\ENDWHILE
\end{algorithmic}
%\vspace{-1mm}
\end{algorithm}

\section{Experimental Results}

To test this algorithm, we use the Daubechies-5 wavelet basis in two dimensions over $32\times 32$ size patches with 5 levels of decomposition. This gives a size $1024$ by $1024$ matrix $\Psi$. We divide this matrix into $49$ sets: $1$ set of the coarsest scale coefficients in a block of size $4\times 4$ containing the two coarsest scales, and $48$ other sets rooted at the coefficients at the next finer scale. Each of these sets contains $21$ $(1+4+16)$ coefficients in a quadtree structure. To create matrix $\Phi$ we first apply a blurring filter $H$ with a symmetric $5\times 5$ kernel that is close to a delta function. This simulates practical optical sampling acquisition effects such as diffraction and helps prevent rank deficiency problems when carrying out inversion. We use different 2D sampling patterns to carry out the subsampling operation represented by matrix S. Hence the acquisition process is represented by $y=\Phi c$ where $\Phi=SH\Psi$. 
The sampling patterns are shown in Table~\ref{Table:Patterns} for each sampling rate $\delta=\frac{M}{N}$ used to generate the results. Each pattern is replicated 8 times in horizontal and vertical directions to give the $32\times 32$ sampling pattern used for matrix $S$. The filter kernel is a $5\times 5$ kernel with $0.29$ at the center and $0.02$ in other locations. The signals are generated by uniformly selecting at random wavelet trees to make the sparsity of the signal the specified value.  The coefficients in these trees are set to values chosen from a standard normal distribution, and the rest are set to zero.

\begin{table}[htbp]

\centering
\subfloat[$\delta=\frac{2}{16}$] {
\begin{tabular}{|c|c|c|c|}
\hline
0 & 0 & 0 & 0 \\ \hline
0 & 1 & 0 & 0 \\ \hline
0 & 0 & 0 & 0 \\ \hline
0 & 0 & 0 & 1 \\ \hline
\end{tabular}}
\qquad
\subfloat[$\delta=\frac{4}{16}$] {
\begin{tabular}{|c|c|c|c|}
\hline
1 & 0 & 0 & 0 \\ \hline
0 & 0 & 1 & 0 \\ \hline
0 & 1 & 0 & 0 \\ \hline
0 & 0 & 0 & 1 \\ \hline
\end{tabular}}
\qquad
\subfloat[$\delta=\frac{6}{16}$] {
\begin{tabular}{|c|c|c|c|}
\hline
1 & 0 & 1 & 0 \\ \hline
0 & 1 & 0 & 1 \\ \hline
1 & 0 & 0 & 0 \\ \hline
0 & 0 & 1 & 0 \\ \hline
\end{tabular}}
\qquad
\subfloat[$\delta=\frac{8}{16}$] {
\begin{tabular}{|c|c|c|c|}
\hline
1 & 0 & 1 & 0 \\ \hline
0 & 1 & 0 & 1 \\ \hline
1 & 0 & 1 & 0 \\ \hline
0 & 1 & 0 & 1 \\ \hline
\end{tabular}}
\qquad
\subfloat[$\delta=\frac{10}{16}$] {
\begin{tabular}{|c|c|c|c|}
\hline
0 & 1 & 0 & 1 \\ \hline
1 & 0 & 1 & 0 \\ \hline
0 & 1 & 1 & 1 \\ \hline
1 & 1 & 0 & 1 \\ \hline
\end{tabular}}
\qquad
\subfloat[$\delta=\frac{12}{16}$] {
\begin{tabular}{|c|c|c|c|}
\hline
1 & 1 & 0 & 1 \\ \hline
0 & 1 & 1 & 1 \\ \hline
1 & 1 & 1 & 0 \\ \hline
1 & 0 & 1 & 1 \\ \hline
\end{tabular}}
\qquad
\subfloat[$\delta=\frac{14}{16}$] {
\begin{tabular}{|c|c|c|c|}
\hline
1 & 1 & 1 & 1 \\ \hline
1 & 1 & 0 & 1 \\ \hline
1 & 1 & 1 & 1 \\ \hline
0 & 1 & 1 & 1 \\ \hline
\end{tabular}}
\caption{Sampling Patterns}
\label{Table:Patterns}
\vspace{-1mm}
\end{table}

\begin{figure}[h]
\vspace{-1mm}
   \centering
   \begin{tabular}{@{\hspace{-2mm}}c@{\hspace{-3mm}}c}
   (a) \includegraphics[width=1.6in]{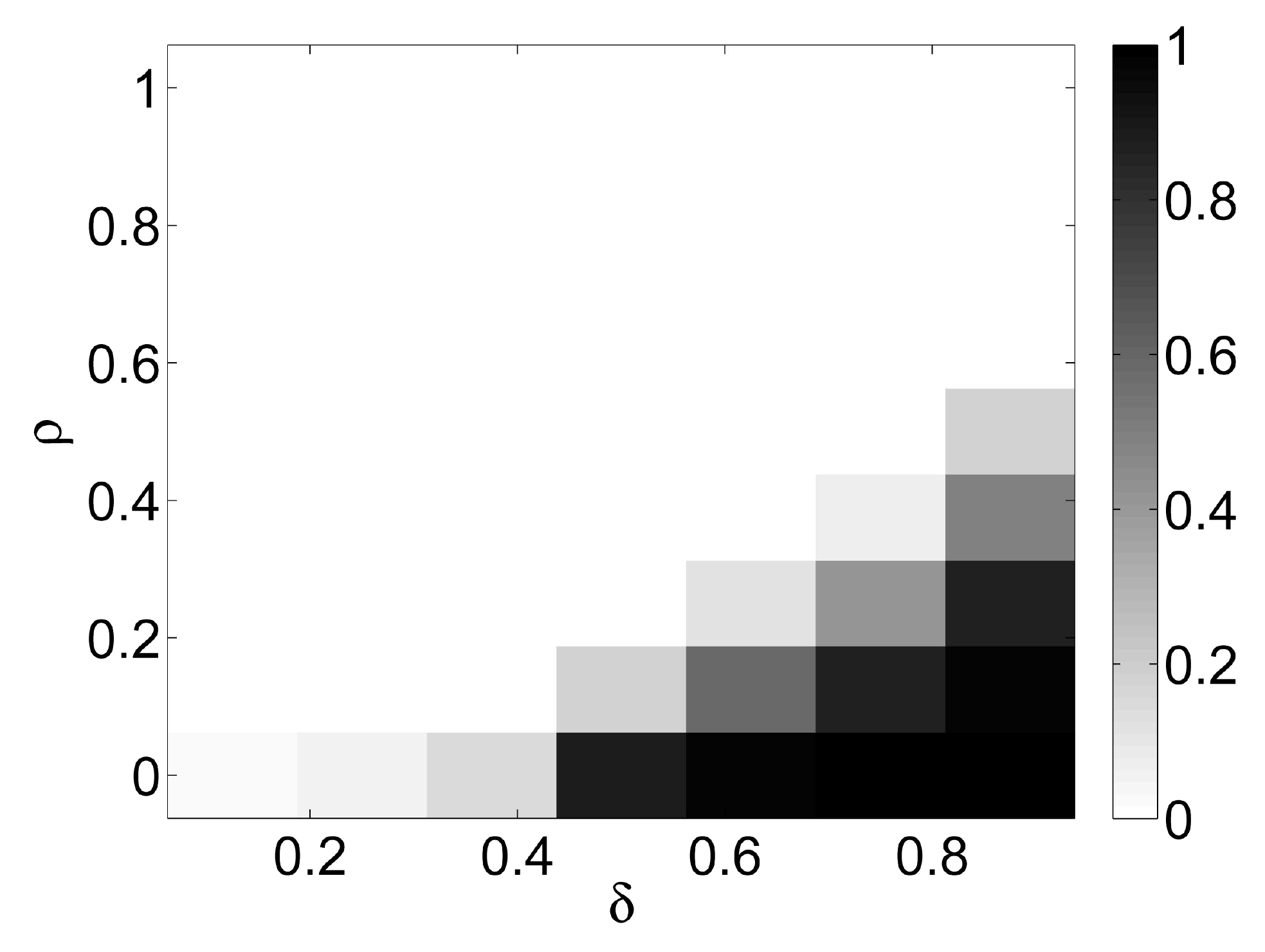} & (b) \includegraphics[width=1.6in]{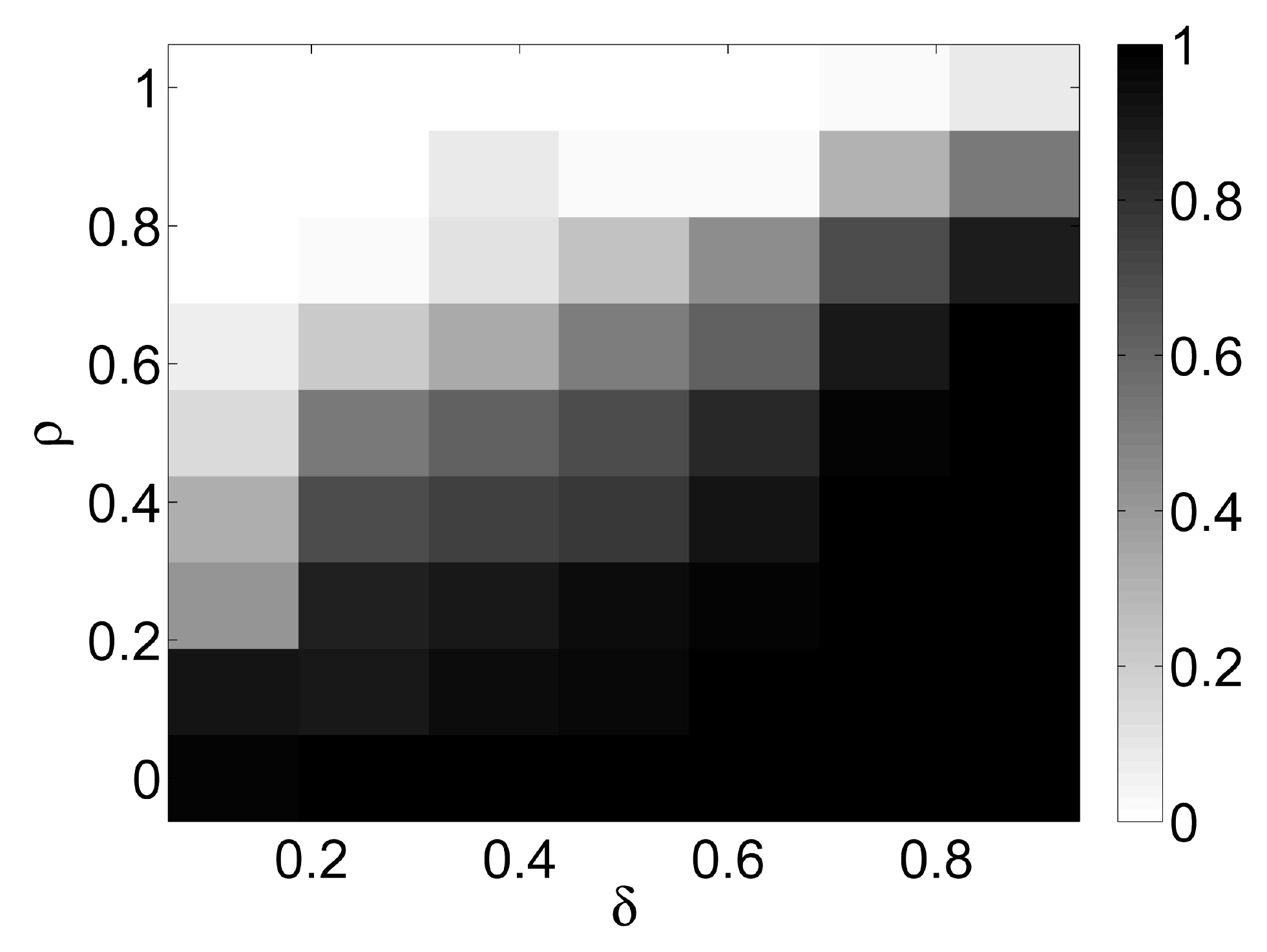}\\
   \end{tabular}
   \vspace{-1mm}
   \caption{\small \sl Proportion of successes with nonzero coefficients concentrated on wavelet trees from (a) $\ell_1$-minimization and (b) PartInv.% for various values of $\delta=\frac{M}{N} \in (0,1)$ (horizontal axis) and $\rho=\frac{K}{M} \in (0,1)$ (vertical axis).
   \label{Table:Wavelet}}
   \vspace{-1mm}
\end{figure}

%\begin{figure}
%\vspace{-1mm}
%   \centering
%\includegraphics[width=2in]{lastplot3.pdf} 
%\vspace{-1mm}
%   \caption{\small \sl Proportion of successes out of 100 trials with nonzero coefficients concentrated on wavelet trees.
%   \label{Table:Wavelet}}
%   \vspace{-1mm}
%\end{figure}

The results are shown in Fig.~\ref{Table:Wavelet}. For each data point we carry out 100 trials.  We declare success if $\frac{1}{N}||c-\hat{c}||^{2}< 10^{-5}$ where $N=32\times 32$.  This shows improvement in selection performance with the sum estimator.  %This estimator is much simpler than the Condensing Sort and Select Algorithm (CSSA) used in \cite{baraniuk2010model} to incorporate wavelet tree structure in CoSAMP but gives better results.  Moreover, CSSA has a complexity of $O(N \log N)$ vs $O(N)$ for the sum estimator, which is called once for every step of the PartInv algorithm. 

\section{Conclusion}

We consider methods of compressive sensing recovery for sampling matrices that have subsets of columns that are strongly intra-correlated, but show low correlation with other subsets. This structure commonly arises in physical sample acquisition/reconstruction scenarios such as image super-resolution. We describe Partial Inversion, an algorithm that improves compressive sensing recovery by removing a source of noise in the initial estimator, and demonstrate its performance by simulations on Gaussian and correlated column subset matrices. We consider compressive sensing recovery when the nonzero coefficients are concentrated on wavelet trees, and demonstrate a simple estimator that improves selection of the trees that carry the nonzero coefficients.

\bibliographystyle{IEEEtran}
\bibliography{cs_struct}

\begin{thebibliography}{10}
\providecommand{\url}[1]{#1}
\csname url@samestyle\endcsname
\providecommand{\newblock}{\relax}
\providecommand{\bibinfo}[2]{#2}
\providecommand{\BIBentrySTDinterwordspacing}{\spaceskip=0pt\relax}
\providecommand{\BIBentryALTinterwordstretchfactor}{4}
\providecommand{\BIBentryALTinterwordspacing}{\spaceskip=\fontdimen2\font plus
\BIBentryALTinterwordstretchfactor\fontdimen3\font minus
  \fontdimen4\font\relax}
\providecommand{\BIBforeignlanguage}[2]{{%
\expandafter\ifx\csname l@#1\endcsname\relax
\typeout{** WARNING: IEEEtran.bst: No hyphenation pattern has been}%
\typeout{** loaded for the language `#1'. Using the pattern for}%
\typeout{** the default language instead.}%
\else
\language=\csname l@#1\endcsname
\fi
#2}}
\providecommand{\BIBdecl}{\relax}
\BIBdecl

\bibitem{superres_survey}
S.~Farsiu, D.~Robinson, M.~Elad, and P.~Milanfar, ``{A}dvances and challenges
  in super-resolution,'' \emph{Int. J. Imag. Syst. Tech.}, vol.~14, no.~2, pp.
  47--57, 2004.

\bibitem{wright_superres}
J.~Yang, J.~Wright, T.~Huang, and Y.~Ma, ``Image superresolution via sparse
  representation,'' \emph{IEEE T. Image Process.}, vol.~19, no.~11, pp.
  2861--2873, Nov. 2010.

\bibitem{candes2006robust}
E.~Cand{\`e}s, J.~Romberg, and T.~Tao, ``Robust uncertainty principles: Exact
  signal reconstruction from highly incomplete frequency information,''
  \emph{IEEE T. Inform. Theory}, vol.~52, no.~2, pp. 489--509, 2006.

\bibitem{candes2006stable}
E.~Candes, J.~Romberg, and T.~Tao, ``Stable signal recovery from incomplete and
  inaccurate measurements,'' \emph{Commun. Pure Appl. Math.}, vol.~59, no.~8,
  pp. 1207--1223, 2006.

\bibitem{donoho1989uncertainty}
D.~Donoho and P.~Stark, ``Uncertainty principles and signal recovery,''
  \emph{SIAM J. Appl. Math.}, vol.~49, no.~3, pp. 906--931, 1989.

\bibitem{candes2011compressed}
E.~Candes, Y.~Eldar, D.~Needell, and P.~Randall, ``Compressed sensing with
  coherent and redundant dictionaries,'' \emph{Appl. Comput. Harmon. A.},
  vol.~31, no.~1, pp. 59--73, 2011.

\bibitem{candes2012towards}
E.~Candes and C.~Fernandez-Granda, ``Towards a mathematical theory of
  super-resolution,'' \emph{Preprint}, 2012.

\bibitem{fannjiang2012coherence}
A.~Fannjiang and W.~Liao, ``Coherence pattern-guided compressive sensing with
  unresolved grids,'' \emph{SIAM J. Imaging Sci.}, vol.~5, no.~1, pp. 179--202,
  2012.

\bibitem{decode_tao}
E.~Candes and T.~Tao, ``Decoding by {L}inear {P}rogramming,'' \emph{IEEE T.
  Inform. Theory}, vol.~51, no.~12, pp. 4203 -- 4215, dec. 2005.

\bibitem{RV08:sparse}
M.~Rudelson and R.~Vershynin, ``On sparse reconstruction from {F}ourier and
  {G}aussian measurements,'' \emph{Comm. Pure Appl. Math.}, vol.~61, pp.
  1025--1045, 2008.

\bibitem{candes_tao}
E.~Candes and T.~Tao, ``Near optimal signal recovery from random projections:
  {U}niversal encoding strategies?'' \emph{IEEE T. Inform. Theory}, vol.~52,
  pp. 5406--5425, Dec. 2006.

\bibitem{candes}
E.~Candes, ``The {R}estricted {I}sometry {P}roperty and its implications for
  {C}ompressed {S}ensing,'' \emph{Cr. Acad. Sci. I-Math.}, pp. 589--592, Dec.
  2008.

\bibitem{CoSaMP}
D.~Needell and J.~Tropp, ``Co{S}a{M}{P}: Iterative signal recovery from
  incomplete and inaccurate samples,'' \emph{Appl. Comput. Harmon. A.},
  vol.~26, pp. 301--321, 2009.

\bibitem{Bjo96:Numerical-Methods}
{\AA}.~Bj{\"o}rck, \emph{Numerical Methods for Least Squares Problems}.\hskip
  1em plus 0.5em minus 0.4em\relax Philadelphia: SIAM, 1996.

\bibitem{l1_magic}
\BIBentryALTinterwordspacing
$\ell_1$-magic. [Online]. Available: \url{http://www.acm.caltech.edu/l1magic/}
\BIBentrySTDinterwordspacing

\end{thebibliography}

\end{document}